\documentclass[preprint,showpacs,preprintnumbers,amsmath,amssymb]{revtex4}

\usepackage{graphicx}
\usepackage{dcolumn}
\usepackage{bm}

\newcommand{\be}{\begin{equation}}  
\newcommand{\ee}{\end{equation}}  
\newcommand{\bear}{\begin{eqnarray}}  
\newcommand{\eear}{\end{eqnarray}}  
\newcommand{\ba}{\begin{array}}  
\newcommand{\ea}{\end{array}}

\newskip\humongous \humongous=0pt plus 1000pt minus 1000pt

\newif\ifdtup

\def\oldreffmt#1{\rlap{[#1]} \hbox to 2\parindent{}}

\def\figfmt#1{\rlap{Figure {#1}} \hbox to 1in{}}  
  
\def\ie{\hbox{\it i.e.}{}}	\def\etc{\hbox{\it etc.}{}}  
\def\eg{\hbox{\it e.g.}{}}

\def\Tr{\mathop{\rm Tr}}

\def\bra#1{\left\langle #1\right|}  
\def\ket#1{\left| #1\right\rangle}

\def\slash#1{#1\!\!\!/\!\,\,}  
\def\beq{\begin{equation}}  
\def\eeq{\end{equation}}  
\def\bea{\begin{eqnarray}}  
\def\eea{\end{eqnarray}}  
\def\half{\frac{1}{2}}  
  
\def\bq{\begin{quote}}  
\def\eq{\end{quote}}

\def\half{\frac{1}{2}}       

\relax  

\newdimen\tdim  
\tdim=\unitlength  
\def\bar{\overline}

\begin{document}


\preprint{FERMILAB-Pub-TM/2341-T}
\title{Lecture Notes for\\
Massless Spinor and Massive Spinor \\
Triangle Diagrams}

\author{Christopher T. Hill}

{\email{hill@fnal.gov}}

\affiliation{
 {{Fermi National Accelerator Laboratory}}\\
{{\it P.O. Box 500, Batavia, Illinois 60510, USA}}
}%

\date{February 1, 2006}

\begin{abstract}
These notes present the details of the computation
of massless and massive spinor
triangle loops for consistent anomalies in gauge theories.
\end{abstract}

\maketitle

\section{Triangle Diagram For Massless Left-Handed spinor}

We 
compute the triangle diagrams and study
the anomalies \cite{jackiw,adler,bardeen,jackiw2} for a pure masssless 
left-handed Weyl spinor in
an external gauge field, with the action:
\beq
S_L = \int d^4x\; \bar{\psi}_L(i\slash{\partial} +\slash{V}_L)\psi_L
\eeq
where:
\beq
V_{L\mu} = B^a_{L\mu} +B^b_{L\mu} + B^c_{L\mu}  
\eeq
couples to the current:
\beq
J_{L\mu} = \bar{\psi}_L \gamma_\mu \psi_L
\eeq
and the components of
$V_L$ have the respective masses $M^{a}$, $M^{b}$, $M^{c}$ \cite{hill}.

This is studied in Section I.A, and the resulting consistent anomalies
are obtained in I.B. These notes are primarily intended to be
a technical memo accompanying \cite{hill}.

In section II we  turn to the case
of a finite, and large electron mass, where ``large''
means in comparison to external momenta and masses.
By expanding in inverse powers of $m^2$ we generate
an operator product expansion whose leading term contains the anomaly.
We carry out the analysis of the loops in
the presence of the full electron mass term, with the couplings
\beq
\label{C3}
\int d^4x\; \left[ \bar{\psi}_L(i\slash{\partial} +\slash{V}_L)\psi_L
+\bar{\psi}_R(i\slash{\partial} +\slash{V}_R)\psi_R
-m(\bar{\psi}_L\psi_R
+ h.c.)\right]
\eeq
where we take separate $L$ and $R$ fields:
\beq
\label{C4}
V_{L\mu} = B^{aL}_\mu + B^{bL}_\mu + B^{cL}_\mu \qquad \qquad
V_{R\mu} = B^{aR}_\mu + B^{bR}_\mu + B^{cR}_\mu  
\eeq
In section 4 we study the anomalies for the left-right symmetric theory
which are used in the main text to obtain the full CS term
physical process of KK-mode decay.
[Note: In the KK-mode case of \cite{hill} we
have: 
\beq
V_{L\mu} = (-1)^a B^a_{\mu} + (-1)^b B^b_{\mu} + (-1)^c B^c_{\mu}  
\eeq 
and 
\beq
V_{R\mu} = B^a_{\mu} + B^b_{\mu} + B^c_{\mu}  
\eeq 
The
phase factors $(-1)^n$ correspond to the conventions for
KK-modes in ref.\cite{hill}, and are constructed so that in
$L$-$R$ symmetric theories, fields
axial vectors ($n$ odd) couple to $\bar{\psi}\gamma_\mu\gamma^5\psi$
with a positive sign. For the purely $LLL$ triangle loops these
factors are irrelevant, and can be absorbed
into the definitions of the fields. In the massive case
where the $R$ and $L$ terms interfere, these signs become relevant.]

While we are computing the triangle loop with
three distinct external fields, $B^a$, $B^b$ and $B^c$, these
can be alternatively viewed as distinct momentum components of the
single field $V$. If all three fields were identical (exact
Bose invariance) the amplitude would vanish, since
it would involve an operator $VVdV$ which is zero. 
It is the external momentum differences 
or flavor indices that distinguish these fields
and allow non-zero operators such as $[B] \sim B^a B^b dB^c$ 
and $[B] \sim B^a B^c dB^b$, \etc. 
In the massless Weyl fermion case of interest presently, we compute
in a limit $M^a >> M^b \sim M^a \sim 0$. We can view 
this as an operator product expansion
of the triangle diagrams in which the internal lines carrying
$p^2 = M_a^2$ are treated as a short-distance expansion.

Both the massless and massive calculations 
confirm Bardeen's result for the consistent anomalies \cite{bardeen} 
and provide
the necessary terms that maintain overall gauge invariance
together with the Chern-Simons term in the process 
$B^a \rightarrow B^b +B^c$, as computed in the text.
 
\begin{figure}[t]  
\vspace{5cm}  
\includegraphics{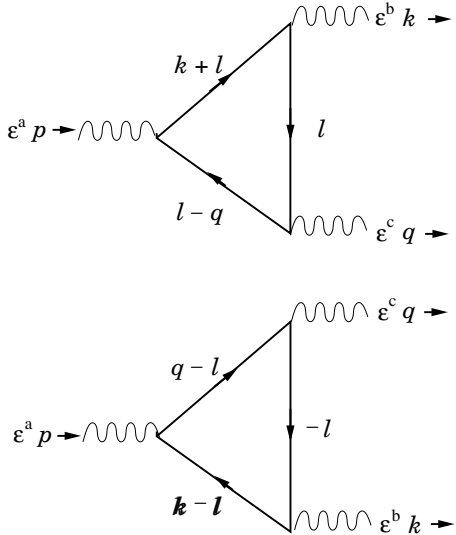}  
\vspace{4.5cm}  
\caption[]{
\addtolength{\baselineskip}{-.3\baselineskip}  
Bose symmetric triangle diagrams 
for $B^a(p)\rightarrow B^b(k)+B^c(q)$
The external lines are on mass-shell, $p^2=M_a^2 $, $k^2=M_b^2$
and $q^2=M_c^2$. The respective polarizations are $\epsilon^a_\mu$,
$\epsilon^b_\mu$ and $\epsilon^c_\mu$. The internal momentum
routing and integration momenta are chosen so that 
both diagrams have a common
denominator. }  
\label{dirac2}  
\end{figure}

\noindent
\subsection{Massless Weyl Spinor}

Our conventions are (Bjorken and Drell, \cite{bj}):
\beq
\epsilon_{0123} = -\epsilon^{0123} = 1
\qquad
\gamma^5 = \gamma_5 = i\gamma^0 \gamma^1 \gamma^2 \gamma^3 =
\left(\begin{array}{cc} 0 & 1 \\ 1 & 0 \end{array}\right) 
\eeq
\beq
1 = \epsilon_{0123} = -\epsilon^{0123}
\eeq
With the particular choice of momentum
routing in the figure, we have the following expression for the sum
of the triangle diagram and its Bose symmetric counterpart,
which have a common denominator:
\bea
 T & = & (-1)(i)^3 (i)^3 \int \frac{d^{4}\ell}{(2\pi)^{4}}
\frac{N_1+N_2}{D} \nonumber \\
N_1 & = & \Tr[\slash{\epsilon}_a L
(\slash{\ell}-\slash{q})
\slash{\epsilon}_c L(\slash{\ell})
\slash{\epsilon}_b L(\slash{\ell}+\slash{k})]
\nonumber \\
N_2 & = & -\Tr[\slash{\epsilon}_a L
(\slash{\ell}+\slash{k})
\slash{\epsilon}_b L(\slash{\ell})
\slash{\epsilon}_c L(\slash{\ell}-\slash{{q}})]
\nonumber \\
D &  = &
(\ell+k)^2(\ell^2)(\ell-q)^2
\eea
where:
\beq
p=k+q;\qquad L =\half(1-\gamma^5)
\eeq
and the overall sign contains $\times(i)^3$ (vertices; note
that our vector potentials have the opposite sign to the
conventions of Bjorken and Drell, hence flipping the vertex
rule from $-i\gamma_\mu \rightarrow +i\gamma_\mu$)), $\times (i)^3$
(propagators), $\times (-1)$ (Fermi statistics).
In $N_2$ we've factored out an overall minus sign.
Note that one must use extreme care to write the 
given correct cyclic ordering of
the factors that make up the numerator, relative to
the momentum routing signs. This affects the overall sign of 
the triangle loop with three gauge vertices (but is has
no effect upon the $im\bar{\psi}\gamma^5\psi $ loop computed in the
massive case).

[For an alternate cyclic ordering, we have:
\bea
N_1 & = & \Tr[\slash{\epsilon}_a L
(-\slash{\ell}-\slash{k})
\slash{\epsilon}_b L(-\slash{\ell})
\slash{\epsilon}_c L(-\slash{\ell}+\slash{q})]
\nonumber \\
N_2 & = & -\Tr[\slash{\epsilon}_a L(-\slash{\ell}+\slash{q})
\slash{\epsilon}_c L(-\slash{\ell})
\slash{\epsilon}_b L(-\slash{\ell}-\slash{{k}})]
\eea
and the momentum factors
appear with minus signs in the numerators. To
understand this, consider the $a$ vertex to be located at position
$x$ in configuration space, for the first Feynman diagram we
have an incoming momentum $p$, 
or an $\exp(-ip\cdot x)$ photon wave-function.
If we suppose the $b$ vertex is located at position $y$.
Then the $k +\ell$ line arises from the 
propagator, $\bra{0}T\psi(x)\bar{\psi}(y)\ket{0}$,
which has a standard Fourier representation of (see, \cite{bj},
Vol. II, p 185):
\beq
i\int \frac{d^4 h}{(2\pi)^4} e^{-ih\cdot(x-y)}\frac{\slash{k}+m}{k^2-m^2}
\eeq
Notice the $\exp(-ih\cdot x)$  factor represents momentum
flowing {\em into} vertex $a$ from vertex $b$.
Likewise, with vertex $c$ located at $z$, 
the $\ell-q$ line has an $\exp(+ih'\cdot (x-z))$, factor 
arising from $\bra{0}T\psi(z)\bar{\psi}(x)\ket{0}$, represents
momentum {\em outflowing} from vertex $a$ toward vertex $c$. 
When we perform the $\int d^4 x A_\mu \bar{\psi}\gamma^\mu \psi$, 
the momentum in the propagators
will satisfy $p+h-h'=0$, and we thus choose to define  $h=-\ell-k$
and $h'=-\ell+q$. ]

We unify the denominator using:
\beq
\frac{1}{ABC} = 2\int_0^1dy\int_0^y dz \frac{1}{(Az + B(y-z)+C(1-y))^3}
\eeq
The unified denominator becomes:
\beq
\frac{1}{D} = 2\int_0^1dy\int_0^y dz 
\frac{1}{( \ell^2 +2\ell\cdot(zk-(1-y)q)+zk^2 +(1-y)q^2  )^3}
\eeq
Shifting the loop momentum to a symmetric
integration momenta, $\overline{\ell}$:
\bea
 \ell & = & \overline{\ell} - zk+(1-y)q
\eea
the unified denominator becomes:
\bea
&  & (\overline{\ell}^2 +z(1-z)k^2 +y(1-y)q^2 
+2k\cdot qz(1-y))
\eea
We define the following vertex tensors :
\bea
\label{match}
A & = & \epsilon_{\mu\nu\rho\sigma}{\epsilon}^\mu_a{\epsilon}^\nu_b
{\epsilon}^\rho_c k^\sigma \qquad \longleftrightarrow \qquad -i \bra{b,k;c,q } 
\epsilon_{\mu\nu\rho\sigma}B^{a\mu} B^{c\nu}\partial^\rho B^{b\sigma}\ket{
a,p }
\nonumber \\
B & = & \epsilon_{\mu\nu\rho\sigma}{\epsilon}^\mu_a{\epsilon}^\nu_b
{\epsilon}^\rho_c q^\sigma  \qquad \longleftrightarrow \qquad 
i \bra{ b,k;c,q  } \epsilon_{\mu\nu\rho\sigma}B^{a\mu}B^{b\nu}
\partial^\rho  B^{c\sigma} \ket{a,p }
\nonumber \\
C & = & \epsilon_{\mu\nu\rho\sigma}{\epsilon}^\mu_a{\epsilon}^\nu_b
{k}^\rho q^\sigma  \qquad \longleftrightarrow  \qquad
\half \bra{ b,k  } F_{\mu\nu}^a \tilde{F}^{b\;\mu\nu} \ket{a,p }
\nonumber \\
D & = & \epsilon_{\mu\nu\rho\sigma}{\epsilon}^\mu_a{\epsilon}^\nu_c
{k}^\rho q^\sigma \qquad \longleftrightarrow  \qquad
-\half \bra{ c,q  } F_{\mu\nu}^a \tilde{F}^{c\;\mu\nu} \ket{a,p }
\nonumber \\
E & = & \epsilon_{\mu\nu\rho\sigma}{\epsilon}^\mu_b{\epsilon}^\nu_c
{k}^\rho q^\sigma  \qquad \longleftrightarrow  \qquad
\half \bra{ b,k;c,q  } F_{\mu\nu}^b \tilde{F}^{c\;\mu\nu} \ket{ 0}
\eea
where we have indicated the corresponding 
operator matrix elements, ( $\bra{out }{\cal{O}}\ket{in} $)
and note:
\beq
\tilde{F}_{\mu\nu} = \half \epsilon_{\mu\nu\rho\sigma}F^{\rho\sigma}
\eeq
is the standard definition of the dual field strength.

\noindent
\subsection{Massless Weyl Spinor Triangle Loops}

We now compute the triangle loops.  Since we are mainly
interested in a heavy KK mode decaying to low mass 
KK-modes, kinematically we have:
\beq
p = k+q \qquad M_b^2 = k^2 \approx 0 \qquad M_c^2 = q^2 \approx 0 \qquad
M_a^2 \approx 2k\cdot q 
\eeq
Hence the large $M_a^2$ limit corresponds to a symmetrical expansion
in $k^2/2k\cdot q $ and $q^2/2k\cdot q$.
We 
have, expanding in $q^2$ and $k^2$, the 
following schematic structure:
\bea
\label{C14}
 T & = & \int' \int \frac{d^{4}\bar\ell}{(2\pi)^{4}}
\frac{\alpha(k,q)\bar\ell^2 + \beta(k,q)}{
(\bar\ell^2 + \Delta^2(k^2,k\cdot q,q^2) )^3} 
\nonumber \\
& = &  \int'\int \frac{d^{4}\bar\ell}{(2\pi)^{4}}\left[
\frac{(\alpha_0\bar\ell^2 + \beta_0
)}{
(\bar\ell^2 + \Delta^2_0 )^3}
\left(1 - \frac{3\gamma q^2 }{(\bar\ell^2 +\Delta^2_0)} 
- \frac{3\delta k^2 }{(\bar\ell^2 +\Delta^2_0)} \right)
+
\frac{\beta_1q^2 + \beta_2k^2 }{
(\bar\ell^2 + \Delta^2_0 )^3}
\right]
\eea
where:
\bea
\gamma & = & y(1-y), \qquad \qquad \delta = z(1-z),\nonumber \\
\Delta^2_0 & = & 2k\cdot q z(1-y)= M_a^2 z(1-y)\; ,
\qquad \qquad
 \int' (\; )  =  2\int_0^1\int_0^y dz \; dy (\;),
\eea
and the $\beta_i$ will be determined below.

For the large mass limit, $|\Delta^2| <<m^2$ 
we define the loop integrals with the usual Wick rotation
on the loop energy 
$\ell_0$ and a Euclidean momentum space cut-of $\Lambda^2$:
\bea
\int \frac{d^4\ell}{(2\pi)^4}\frac{(1,\;\ell^2)}{(\ell^2 -m^2+i\epsilon)^3}
& = & \left[ \frac{-i}{16\pi^2}\left(  \frac{1}{2m^2 }\right), 
\frac{i}{16\pi^2}\left[\ln\left(\frac{\Lambda^2}{m^2}\right) 
- \frac{3}{2}\right]\right]
\nonumber \\
\int \frac{d^4\ell}{(2\pi)^4}\frac{(1,\;\ell^2)}{(\ell^2 -m^2+i\epsilon)^4}
& = &
\left[ \frac{i}{16\pi^2}\left(  \frac{1}{6(m^2)^2 }\right),
 \frac{-i}{16\pi^2}\left(  \frac{1}{3(m^2)^2 }\right)\right]
\eea
The familiar Wick rotation is a counterclockwise rotation of the contour
of the  $\ell_0$ integral in the complex plane. The rotation is
clockwise to avoid the poles at $\pm\sqrt{\vec{l}^2+m^2}\mp i\epsilon$.
The new contour in $\ell_0$ runs from $-i\infty $ to $+i\infty$,
and we rescale to a Euclidean momentum 
$\ell_0 = i\bar{\ell}_0$ where  $d^4\ell \rightarrow
id^4\bar{\ell}$ 
Since the results are analytic functions of
$m^2$, in the limit $|\Delta_0^2| >> m^2$
we simply replace $m^2\rightarrow -\Delta_0^2$:

We thus have the result: 
\bea
T & = & \frac{i}{16\pi^{2}}
 \int'\left[{\alpha_0}\left( \ln\left(\frac{\Lambda^2}{-\Delta^2_0}\right) 
 -\frac{3}{2} 
+\frac{\gamma q^2}{\Delta^2_0} +\frac{\delta k^2}{\Delta^2_0}\right)\right.
\nonumber \\
& & \left.
+ \beta_0 \left( \frac{1}{2\Delta^2_0} -\frac{\gamma q^2}{2(\Delta_0^2)^2}
-\frac{\delta k^2}{2(\Delta_0^2)^2}  \right)
+ \frac{\beta_1q^2}{2\Delta^2_0} + \frac{\beta_2k^2}{2\Delta^2_0} \right]
\eea
We turn to compute $\alpha_0$ and the $\beta_i$.
The numerators are expressed in terms of the shifted momenta
and we keep only the $\gamma^5$ term in the $L$ projection:
\bea
N_1 & = & +\half\Tr[\gamma^5\slash{\epsilon}_a
(\slash{\bar{\ell}}-z\slash{k}-y\slash{q})
\slash{\epsilon}_c
(\slash{\bar{\ell}}-z\slash{k}+(1-y)\slash{q})
\slash{\epsilon}_b (\slash{\overline{\ell}}+(1-z)\slash{k}+(1-y)\slash{q})
]
\nonumber \\
N_2 & = & -\half\Tr[\gamma^5\slash{\epsilon}_a
(\slash{\bar\ell}+(1-z)\slash{k}+(1-y)\slash{q})
\slash{\epsilon}_b 
(\slash{\bar\ell}-z\slash{k}+(1-y)\slash{q})
\slash{\epsilon}_c 
(\slash{\bar\ell}-z\slash{k}-y\slash{q})
]
\nonumber \\
\eea
Using the identities:
\bea
\slash{a}\slash{b}\slash{c} & = &
a\cdot b \slash{c} - a\cdot c \slash{b} + b\cdot c \slash{a}
+i\epsilon_{\mu\nu\rho\sigma} a^\mu b^\nu c^\rho \gamma^5\gamma^\sigma
\nonumber \\
\slash{a}\slash{b}\slash{c}-\slash{c}\slash{b}\slash{a} & = &
2i\epsilon_{\mu\nu\rho\sigma} a^\mu b^\nu c^\rho 
\gamma^5\gamma^\sigma
\nonumber \\
\Tr(\slash{a}\slash{b}\slash{c}\slash{d})
 & = & 4i\epsilon_{\mu\nu\rho\sigma} a^\mu b^\nu c^\rho d^\sigma
 \nonumber \\
 \gamma_\mu \gamma^\nu \gamma^\mu & = & -2
\gamma^\nu \qquad \qquad 
 \gamma_\mu \gamma^\rho \gamma^\nu \gamma^\sigma \gamma^\mu  = 
 -2\gamma^\sigma \gamma^\nu \gamma^\rho
\eea
we first compute the $\bar{\ell}^2$ terms, using $\ell_\mu\ell_\nu
\rightarrow g_{\mu\nu}\ell^2/4$. 
From the 
expansion in $\bar{\ell}^2$ of eq.(\ref{C14}), for the numerator terms
we have:
\beq
N_1+N_2 \equiv \alpha(k,q)\bar{\ell}^2 + \beta(k,q)
\eeq
We obtain the leading terms:
\bea
\alpha(k,q) & = & 
-\; \frac{1}{4}\overline{\ell}^2\Tr[\gamma^5\slash{\epsilon}_a
\slash{\epsilon}_c 
\slash{\epsilon}_b ((1-z)\slash{k}+(1-y)\slash{q})
+
\gamma^5\slash{\epsilon}_a
(-z\slash{k}-y\slash{q})
\slash{\epsilon}_c 
\slash{\epsilon}_b 
\nonumber \\
& & \qquad \qquad \qquad
+
\gamma^5\slash{\epsilon}_a
\slash{\epsilon}_b 
(-z\slash{k}+(1-y)\slash{q})
\slash{\epsilon}_c 
]
\nonumber \\
& &
\;+ \frac{1}{4}\overline{\ell}^2
\Tr[\gamma^5\slash{\epsilon}_a
\slash{\epsilon}_b
\slash{\epsilon}_c (-z\slash{k}-y\slash{q})
+
\gamma^5\slash{\epsilon}_a(
(1-z)\slash{k}+(1-y)\slash{q})
\slash{\epsilon}_b
\slash{\epsilon}_c 
\nonumber \\
& & \qquad \qquad \qquad
+
\gamma^5\slash{\epsilon}_a
\slash{\epsilon}_c
(-z\slash{k}+(1-y)\slash{q})
\slash{\epsilon}_b 
]
\nonumber \\
 & = & 2i\overline{\ell}^2 \big[
(1-3z)[\epsilon_{\mu\nu\rho\sigma}{\epsilon}^\mu_a{\epsilon}^\nu_b
{\epsilon}^\rho_c k^\sigma ]
+
(2-3y)
[\epsilon_{\mu\nu\rho\sigma}{\epsilon}^\mu_a{\epsilon}^\nu_b
{\epsilon}^\rho_c q^\sigma ]
\big]
\eea
whence:
\bea
 \alpha_0 & = & 2i\big[
(1-3z)A +(2-3y)B 
\big]
\eea
We can see that $\alpha_0$ is Bose symmetric under $b\leftrightarrow \gamma$,
$k\leftrightarrow q$. 

We also have:
\bea
\beta  & = & - \half\Tr\big[\gamma^5 
[(1-z)\slash{k}+(1-y)\slash{q}]\big]\slash{\epsilon}_a[
z\slash{k}+y\slash{q}]
\slash{\epsilon}_c 
[z\slash{k}-(1-y)\slash{q}]\slash{\epsilon}_b\big]
\nonumber \\
& & 
+ \half\Tr\big[\gamma^5[z\slash{k}+y\slash{q}]\slash{\epsilon}_a[
(1-z)\slash{k}+(1-y)\slash{q}]
\slash{\epsilon}_b
[z\slash{k}-(1-y)\slash{q}]
\slash{\epsilon}_c \big]
\eea
Note that here we have rearranged the order of terms in $N^{(0)}_2$
in a manner that preserves the manifest Bose
symmetry, $c\leftrightarrow b$, $k\leftrightarrow q$.
$\beta$ is readily evaluated:
\bea
\beta & = & 
\; +4i\epsilon_{\mu\nu\rho\sigma}
[z{k}^\mu + y{q}^\mu]
{\epsilon}^\nu_a [(1-z){k}^\rho+(1-y){q}^\rho]\times
\nonumber \\
& & \;\;\;\;\;
\big[{\epsilon}^\sigma_c
{\epsilon}_b\cdot[
(z{k}-(1-y){q}]
+
{\epsilon}^\sigma_b[
(z{k}-(1-y){q}]\cdot{\epsilon}_c
-[z{k}^\sigma-(1-y){q}^\sigma]{\epsilon}_b\cdot\epsilon_c \big]
\nonumber \\ 
& & 
-4i\epsilon_{\mu\nu\rho\sigma}\epsilon^\mu_b[
z{k}^\nu-(1-y){q}^\nu]
{\epsilon}^\rho_c 
\times\nonumber \\ & & 
\left( [z{k}^\sigma+y{q}^\sigma]
[(1-z){k}\cdot{\epsilon_a}+(1-y){q}\cdot{\epsilon_a}]
+
[(1-z){k}^\sigma+(1-y){q}^\sigma]
[z{k}\cdot{\epsilon_a}+y{q}\cdot{\epsilon_a}] \right.
\nonumber \\
& & 
\left.\qquad \qquad \qquad \qquad \qquad \qquad 
-\epsilon^\sigma_a [z(1-z){k}^2+(z+y-2yz)k\cdot{q}+y(1-y)q^2]
\right)
\nonumber \\
& = & 4i[A(z)-B(1-y)]\big(z(1-z)k^2+(z+y-2yz)k\cdot q +y(1-y)q^2 \big)
\nonumber \\
& & 
+4iE\big[ k \cdot \epsilon_a (2z-zy-z^2)
+q\cdot\epsilon_a (z+y-zy-y^2)\big]
\nonumber \\
& & 
-4i[D][z(z-y)\epsilon_b\cdot k-(z-y)(1-y)\epsilon_b\cdot q]
\nonumber \\
& & 
-4i[C][z(z-y)\epsilon_c\cdot k-(z-y)(1-y)\epsilon_c\cdot q]
\eea
Note that Bose symmetry is manifest
in the $[E]$ term since
 under the replacement, $z\leftrightarrow 1-y$
we have that $(2z -zy -z^2)\leftrightarrow (z+y-zy-y^2)$.
Also, $z^2-zy\leftrightarrow z-y+y^2-zy$, and $C\leftrightarrow -D$,
$\epsilon_b\leftrightarrow \epsilon_c$
makes the symmetry manifest in last two terms.

It is useful to rearrange the $\beta$ term to
reflect the $b$ and $c$ gauge invariance in the $k^2=q^2=0$ limit:
\bea
\beta & = & 4i[A]\big(2k\cdot q z^2(1-y)  +q^2 (1-y)(y-z+zy) + k^2z^2(1-z)\big)
\nonumber \\
& & 
-4i[B]\big(2k\cdot q z(1-y)^2 + q^2y(1-y)^2 + k^2z(1-2z+yz) \big)
\nonumber \\
& & 
-4i[C k \cdot \epsilon_c +A k \cdot q]z(z-y)
\nonumber \\
& & 
+4i[C  q\cdot\epsilon_c + Aq^2] (z-y) (1-y)
\nonumber \\ & & 
-4i[D\epsilon_b\cdot k+  Bk^2] z(z-y)
\nonumber \\ & & 
+4i[D\epsilon_b\cdot q +B k\cdot q](z-y)(1-y)
\nonumber \\ & & 
+4i[E\epsilon_a\cdot k](2z-zy-z^2)
+4i[E\epsilon_a\cdot q](z+y-zy-y^2)
\eea
Here the third through last lines are arranged to vanish
if $\epsilon^b_\mu \rightarrow k_\mu$ 
and $\epsilon^c_\mu \rightarrow q_\mu$ since $A\rightarrow -C$ and
$B\rightarrow -D$, and $E\rightarrow 0$. 
It can be readily checked that this expression is Bose symmetric.
We can therefore extract the $\beta_i$:
\bea
\label{C30}
\beta_0 & = & 
4i[A]\big(2k\cdot q \; z^2(1-y)\big)
\nonumber \\
& & 
-4i[B]\big(2k\cdot q \; z(1-y)^2 \big)
\nonumber \\
& & 
-4i[C k \cdot \epsilon_c +A k \cdot q]z(z-y)
\nonumber \\
& & 
+4i[C  q\cdot\epsilon_c + Aq^2] (z-y) (1-y)
\nonumber \\ & & 
-4i[D\epsilon_b\cdot k+  Bk^2] z(z-y)
\nonumber \\ & & 
+4i[D\epsilon_b\cdot q +B k\cdot q](z-y)(1-y)
\nonumber \\ & & 
+4i[E\epsilon_a\cdot k](2z-zy-z^2)
+4i[E\epsilon_a\cdot q](z+y-zy-y^2)
\nonumber \\
\beta_1 & = & +4i[A](1-y)(z-y+zy)  -  4i[B]y(1-y)^2 
\nonumber \\
\beta_2 & = & +4i[A]z^2(1-z)   -    4i[B]z(1-2y+yz)  
\eea

We now assemble the result. 
The superficially log divergent with $k^2=q^2=0$ (the $\alpha_0$ 
contribution)
yields a finite result:
\bea
\label{an1}
T_0 
& = &
 \frac{2i}{16\pi^2}\;\int_0^1dy\int_0^y dz\;
(2i)\big[(1-3z)A + (2-3y)B \big]\left[\ln\left(\frac{\Lambda^2}{-z(1-y)M_a^2 }\right) 
-\frac{3}{2} \right]
\nonumber \\
& = &
-\frac{1}{24\pi^2}[A]+\frac{1}{24\pi^2}[B]
\eea
$T_0$ is Bose symmetric under interchange of
the photon and $b$-KK mode (let $q\leftrightarrow k$, hence $M_b^2\rightarrow
0$,
$\epsilon_a\leftrightarrow
\epsilon_b$, and note $A\leftrightarrow B$).  Note, as a check on
the large $m^2$ case, that if the argument of
the log, ${\Lambda^2}/{z(1-y)M_a^2 }$ is replaced by ${\Lambda^2}/{m^2}$
then $T_0 = 0$. The result is finite because the
integrals:
\bea
\label{an2}
\int_0^1dy\int_0^y dz\;
\big[(1-3z)A + (2-3y)B \big]
& = & 0
\eea
This furthermore implies that the imaginary part of
the expression is vanishing.

The remaining terms of the triangle diagrams 
to order  $q^2$, $k^2$ are:
\bea
T_1 & = & \frac{2i}{16\pi^2}\int_0^1dy\int_0^y dz\; 
\left(
\frac{\beta_0}{2\Delta^2_0} -\frac{\beta_0\gamma q^2}{2(\Delta_0^2)^2}  
+ \frac{\alpha_0 \gamma q^2}{\Delta_0^2} +\frac{\beta_1q^2}{2\Delta_0^2}
\right)
\eea
Combining all of
the above terms yields the following full result for the triangle
diagrams:
\bea
T_{0} + T_{1} & = & 
-\frac{1}{12\pi^2}[A] + \frac{1}{12\pi^2}[B]
\nonumber \\
& & +\frac{1}{4\pi^2}[Ck\cdot \epsilon_c  + Ak\cdot q]\frac{I_b}{M_a^2}
\nonumber \\
& & -\frac{1}{4\pi^2}[C q\cdot \epsilon_c  + Aq^2]\frac{I_c}{M_a^2}
\nonumber  \\
& & +\frac{1}{4\pi^2}[D\epsilon_b\cdot k  + B k^2 ]\frac{I_b}{M_a^2}
\nonumber  \\
& & -\frac{1}{4\pi^2}[D\epsilon_b\cdot q  + B k\cdot q ]\frac{I_c}{M_a^2}
\nonumber  \\
& &
-\frac{1}{4\pi^2\; M_a^2}[E](\epsilon_a\cdot k \; I'_b + \epsilon_a\cdot q \; I'_c)
+ {\cal{O}}(q^2,k^2).
\eea
The integrals
$I_i$
and $I'_i$ are infrared divergent in our expansion. The result
is manifestly Bose symmetric only if we perform the unification
integrals, $I_i$ and $I'_i$ with 
a Bose symmetric IR cut-off. For a particular choice of small 
IR cut-offs
$x_i$, the leading log divergent terms are:
\bea
I_b & = & \int_0^{1-x_b}\int_0^y dz\;dy\;\frac{z(z-y)}{z(1-y)} 
= \half \ln(x_b) + k_b
\nonumber \\
I_c & = & \int_0^1\int_x^y dz\;dy\;\frac{(1-y)(z-y)}{z(1-y)} 
=\half \ln(x_c) + k_c
\nonumber \\
I'_{b}& = &\int_0^{1-x}\int_0^y dz\;dy\;\frac{2z-zy-z^2}{z(1-y)} = - \half \ln(x_b)
+k'_b
\nonumber \\
I'_{c}& = & \int_0^1\int_x^y dz\;dy\;\frac{z+y-zy-y^2}{z(1-y)} = - \half \ln(x_c)
+k'_c
\eea
Note that $I_b=I_c$ and $I'_b=I'_c$ 
if $x_b = x_c$ and Bose symmetry is maintained.
The physical cutoffs are of order 
$x_b \sim M_b^2/M_a^2$ and $x_c \sim M_c^2/M_a^2$,
and their exact  coefficients are indeterminate in our expansion
(hence finite corrections, $k$ and $k'$ to the logs are indeterminate).
This can presumably be replaced with a more physical procedure by
resumming $k^2$ and $q^2$ into the denominators.
The logarithmic IR singularities in, \eg, the $ q^2 =0$
limit are presumably cancelled by collinear $\bar{\psi}\psi $ 
propagation in the $B^c\rightarrow B^b+ \bar{\psi}+\psi$
process, where $\bar{\psi}+\psi$ rescatter into a photon.

Note a final lemma that is relevant to the anomaly:
\beq
\label{sumI}
I_a + I_b + I'_b + I'_c = 2
\eeq
which is an infra-red ``safe'' quantity.

\subsection{Massless Weyl Spinor Anomaly}

The amplitude $T$ that we have just computed is:
\beq
T 
= \bra{b,c}T ...\; 
i\int d^4 x\; \exp(-ip\cdot x)\epsilon^{a}_{\mu}\bar{\psi}\gamma ^\mu
\psi_L \; ... \ket{0}
\eeq
On the other hand, the amplitude we want is the matrix
element of the current divergence:
\bea
W  & = & \bra{b,c}T ... 
\int d^4 x\; \exp(-ip\cdot x)\partial_{\mu}\bar{\psi}\gamma ^\mu
\psi_L \; ... \ket{0} \nonumber \\
& = & \bra{b,c}T ... 
\int d^4 x\; (-\partial_{\mu}\exp(-ip\cdot x)) \bar{\psi}\gamma ^\mu \psi_L
\; ... \ket{0}
\nonumber \\
& = & \bra{b,c}T ... 
\int d^4 x\; \exp(-ip\cdot x) ip_\mu \bar{\psi}\gamma ^\mu \psi_L
\; ... \ket{0}
\eea
We thus obtain $W$ from $T$ by the replacement:
\beq
W = T(\epsilon^{a}_{\mu}\rightarrow p_\mu)
\eeq
Under this substitution we have:
$[A]\rightarrow -[E]$, $[B]\rightarrow [E]$,
$[C]\rightarrow 0$, $[D]\rightarrow 0$,  
and  $\epsilon_a\cdot k \rightarrow k\cdot q$, 
$\epsilon_a\cdot q \rightarrow k\cdot q$.

We thus obtain:
\bea
T_{0}+T_{1} & \rightarrow  & 
\frac{1}{12\pi^2}[E] + \frac{1}{12\pi^2}[E] -
\frac{1}{8\pi^2}(I_a + I_b + I'_b + I'_c)[E]
\nonumber \\
& = & -\frac{1}{12\pi^2}[E]\qquad 
\eea
where we use eq.(\ref{sumI}).
The result is infra red non singular. 

Note that we have the operator correspondence
$[E]\rightarrow (1/4)F\tilde{F}$.
Our result for the anomaly thus
corresponds to the operator equation:
\beq
\partial^\mu \bar{\psi}\gamma_\mu\psi_L =
-\frac{1}{48\pi^2}F_{L\mu\nu}\tilde{F}^{\mu\nu}_L 
\eeq
which agrees 
with Bardeen's result for the left-right symmetric
anomaly in the case of a massless Weyl spinor \cite{bardeen}.

As a further check on the calculation, 
we can also examine the anomaly in the $B^c$ current, 
by letting 
$\epsilon_c\rightarrow -q$ (the minus
sign occurs since $B^c$ is outgoing), and we take the $c$ field to be
 on-shell and massless, \ie.
set $q^2=0$. 
 Whence $[A] =[C]$ and $[B]=[D]=[E] =0$:
\bea
T_0+T_1 & \rightarrow &  -\frac{1}{12\pi^2} [C] 
\eea
Using eq.(\ref{match}) this corresponds to:
\beq
\partial^\mu \bar{\psi}\gamma_\mu\psi_L =
-\frac{1}{48\pi^2}F_{L\mu\nu}\tilde{F}^{\mu\nu}_L \leftrightarrow 
-\frac{1}{24\pi^2}F_{a\mu\nu}\tilde{F}^{\mu\nu}_b
\eeq
consistent with the $a$ channel result. 
Likewise, we can check the $B^b$ channel, and 
verify the same result.

We can furthermore check the
off-shell gauge invariance for $c$ identified
with a photon and $M_c^2 =0$. We again
set $\epsilon_c\rightarrow -q$ and examine
the  ${\cal{O}}(q^2)$ terms.  Whence $[A] = [C]$ and $[B]=[D]=[E] =0$:
\bea
-\frac{\beta_0\gamma q^2}{2\Delta_0^4} & \rightarrow &
2iCq^2\left(\frac{zy(1-y)}{z(1-z)k^2+2k\cdot q z(1-y)}  \right)
\nonumber \\
\frac{\beta_1 q^2}{2\Delta_0^2} &\rightarrow &
2iCq^2\left(\frac{(1-y)(z-y-zy)}{z(1-z)k^2 +2k\cdot q z(1-y)}  \right)
\nonumber \\
\frac{\gamma q^2}{\Delta_0^2}\alpha_0 & \rightarrow &
2iCq^2\left(\frac{(1-3z)y(1-y)}{z(1-z)k^2 +2k\cdot q z(1-y)}  \right)
\eea
The sum of these terms in the amplitude is:
\bea
T_1\;q^2 & \rightarrow & \frac{i}{16\pi^2}(2i [C]q^2)\int_0^1dy\int_0^y dz
\left(
\frac{z(1-3z)(1-y)q^2}{z(1-z)k^2+2k\cdot q z(1-y)}\right)
\nonumber \\
& = & 0
\eea
since the following integral miraculously vanishes:
\bea
0 & = & \int_0^1dy\int_0^y dz
\frac{(1-3z)(1-y)}{(1-z)X+(1-y)Y}
\eea
Thus the $O(q^2)$ terms in the current divergence
vanish and off-shell gauge invariance is maintained.
The associated operators thus contain internal factors
of $\partial_\mu F^{\mu\nu}$. They can presently be set to zero
by use of equations of motion. More generally they
are easily seen to
have associated log IR divergences in the limit of
zero electron mass, which are presumably
associated with IR singularities in $B^a\rightarrow B^b
+ e^+e^-$ where the massless 
electron pair becomes indistinguishable from
the photon. The IR singularity is cut-off by a
nonzero electron mass.

This implies that the only non-gauge invariant 
part of the amplitude is the
anomaly.

\noindent
\section{Finite Electron Mass}

\subsection{Triangle Loops}

We now turn to the case
of a finite, and large electron mass, where ``large''
means in comparison to external momenta and masses.
By expanding in inverse powers of $m^2$ we generate
an operator product expansion whose leading term contains the anomaly.
We carry out the analysis of the loops in
the presence of the full electron mass term, with the couplings
\beq
\int d^4x\; \left( \bar{\psi}_L(i\slash{\partial} +\slash{V}_L)\psi_L
+\bar{\psi}_R(i\slash{\partial} +\slash{V}_R)\psi_R
-m(\bar{\psi}_L\psi_R
+ h.c.)\right)
\eeq
where we take separate $L$ and $R$ fields,
$B_\mu^{a\;L,R}$:
\beq
V_{L\mu} = B^{aL}_\mu + B^{bL}_\mu + B^{cL}_\mu \qquad \qquad
V_{R\mu} = B^{aR}_\mu + B^{bR}_\mu + B^{cR}_\mu  
\eeq
[Note that in comparison to the KK-mode normalizations used
in \cite{hill} we have:
\beq
B^n_{L\mu} = (-1)^n B^n_{\mu} \qquad \qquad
B^n_{R\mu} = B^n_{\mu} 
\eeq 
We will implement this relationship subsequently, but presently
we work in the independent and generic  $V_L$, $V_R$ basis.]

We presently adopt an obvious generalized notation for vertices, \eg,
\bea
A^{LRL} & = & \epsilon_{\mu\nu\rho\sigma}{\epsilon}^{L\mu}_a{\epsilon}^{R\nu}_b
{\epsilon}^{L\rho}_c k^\sigma \;,
\qquad\qquad 
A^{LRR}  =  \epsilon_{\mu\nu\rho\sigma}{\epsilon}^{L\mu}_a{\epsilon}^{R\nu}_b
{\epsilon}^{R\rho}_c k^\sigma \qquad   .\;.\;. 
\nonumber \\
C^{LR} & = & \epsilon_{\mu\nu\rho\sigma}{\epsilon}^{L\mu}_a{\epsilon}^{R\nu}_b
{k}^\rho q^\sigma  \qquad   .\;.\;. 
\eea
and so forth. 

The $LLL$ 
($RRR$) loops have just been computed, arising from the pure 
massless $\psi_L$  ($\psi_R$). In the case of  a massive
electron 
the $LLL$ ($RRR$) loops have the same
numerator structure, but the  denominator now 
contains  electron mass terms:
\bea
D &  = &
[(\ell+k)-m^2][(\ell^2)^2-m^2][(\ell-q)^2-m^2]
\eea
This causes all of the previously computed $LLL$ 
($RRR$) terms to become suppressed
in the large $m^2$ limit. For example, the $\alpha_0$ term
previously computed for $m^2=0$
now becomes:
\bea
\label{an3}
T_0 
& = &
 -\frac{1}{4\pi^2}\;\int_0^1dy\int_0^y dz\;
\big[(1-3z)A + (2-3y)B \big]\left[\ln\left(\frac{\Lambda^2}{m^2-z(1-y)M_a^2}\right) 
-\frac{3}{2} \right]
\nonumber \\
& \longrightarrow &
-\frac{M_a^2}{480\pi^2 m^2 }\left([A] - [B]\right),
\eea
and now vanishes in the large $m^2$ limit.
All of the new terms of interest in the massive electron
case arise from the numerator terms containing mass insertions.
This represent mixing from $\psi_L$ to the  $\psi_R$,
and thus generates new vertices, such as $[A]^{LRL}$, \etc.  

Let us compute the triangle loops with a
single pure left-handed $\epsilon^{aL}_\mu \gamma^\mu L$ vertex, carrying
in momentum $p$, and again noting the 
the cyclic order in which numerator terms are written:
\bea
 T_L & = & (-1)(i)^3(i)^3\int \frac{d^{4}\ell}{(2\pi)^{4}}
\frac{N_1+N_2}{D} \nonumber \\
N_1 & = &
\Tr[\slash{\epsilon}_a L(\slash{\ell}-\slash{q}+m)
(\slash{\epsilon}^L_c L+\slash{\epsilon}^R_c R) (\slash{\ell}+m)
(\slash{\epsilon}^L_b L+\slash{\epsilon}^R_b R) (\slash{\ell}+\slash{{k}}+m)]
\nonumber \\
N_2 & = &  -\Tr[\slash{\epsilon}_a L(\slash{\ell}+\slash{k}-m)
(\slash{\epsilon}^L_b L+\slash{\epsilon}^R_b R)(\slash{\ell}-m)
(\slash{\epsilon}^L_c L+\slash{\epsilon}^R_c R)(\slash{\ell}+\slash{q}-m)]
\nonumber \\
D &  = &
[(\ell+k)^2-m^2][\ell^2-m^2][(\ell-q)^2-m^2]
\eea
Note the sign flips in the momentum and $m$ terms in $N_1$
and momenta in $N_2$, a consequence
of having factored out an overall minus sign (note: we
could have written a different cyclic order giving
the more conventional signs).
Upon unifying denominator factors and 
shifting the loop momentum as before, we  obtain
the unified denominator:
\beq
( \ell^2 +2\ell\cdot(zk-(1-y)q)+zk^2 +(1-y)q^2 -m^2 )^3
\eeq
The numerators become:
\bea
N_1 & = & 
\Tr[\slash{\epsilon}^L_a L(\slash{\bar\ell}
-z\slash{k}-y\slash{q}+m)
(\slash{\epsilon}^L_c L+\slash{\epsilon}^R_c R) 
\times \nonumber \\ & &
(\slash{\bar\ell}-z\slash{k}+(1-y)\slash{q}+m)
(\slash{\epsilon}^L_b L+\slash{\epsilon}^R_b R) 
(\slash{\bar\ell}+(1-z)\slash{k}+(1-y)\slash{q}+m)]
\nonumber \\
N_2 & = & 
-\Tr[\slash{\epsilon}^L_a L (\slash{\overline{\ell}}+
(1-z)\slash{k}+(1-y)\slash{q}-m)
(\slash{\epsilon}^L_b L+\slash{\epsilon}^R_b R)
\times\nonumber \\ & & 
(\slash{\bar{\ell}}-z\slash{k}+(1-y)\slash{q}-m)
(\slash{\epsilon}^L_c L+\slash{\epsilon}^R_c R) 
(\slash{\bar{\ell}}-z\slash{k}-y\slash{q}-m)]
\eea
Defining the expansion in $\bar{\ell}^2$:
\beq
N_1+N_2 = \alpha_0\bar{\ell}^2 + \beta + m^2 \omega
\eeq
we see that the $\alpha_0$ term is just the previously
computed $LLL$ term, which now produces a result that vanishes in
the large $m^2$ limit as we have just noted above. 
Similarly, the $\beta $ term is as before, 
entirely composed of $LLL$ terms,  and 
reproduces the previously computed terms 
in the massless case, but with $1/M_a^2 $ now replaced
by $1/m^2$. Thus the only new effects 
of interest are contained in the $m^2\omega$ term.

Expanding the numerators with the shifted loop momentum we
have the leading $m^2$ term:
\bea
m^2\omega & = & +\frac{m^2}{2}\Tr[\gamma^5\slash{\epsilon}^L_a
\slash{\epsilon}^R_c 
\slash{\epsilon}^L_b (\slash{\bar\ell}+(1-z)\slash{k}+(1-y)\slash{q})
+
\gamma^5\slash{\epsilon}^L_a(\slash{\bar\ell}
-z\slash{k}-y\slash{q})
\slash{\epsilon}^L_c 
\slash{\epsilon}^R_b 
+\nonumber \\
& & \qquad \qquad 
\gamma^5\slash{\epsilon}^L_a
\slash{\epsilon}^R_c 
(\slash{\bar\ell}-z\slash{k}+(1-y)\slash{q})
\slash{\epsilon}^R_b ]
\nonumber \\
& &
-\frac{m^2}{2}\Tr[\gamma^5\slash{\epsilon}^L_a 
\slash{\epsilon}^R_b
\slash{\epsilon}^L_c (\slash{\bar{\ell}}-z\slash{k}-y\slash{q})
+
\gamma^5\slash{\epsilon}^L_a(\slash{\overline{\ell}}+
(1-z)\slash{k}+(1-y)\slash{q})
\slash{\epsilon}^L_b\slash{\epsilon}^R_c 
+\nonumber \\
& & \qquad \qquad \gamma^5\slash{\epsilon}^L_a
\slash{\epsilon}^R_b
(\slash{\bar{\ell}}-z\slash{k}+(1-y)\slash{q})
\slash{\epsilon}^R_c 
]
\eea
Since $(\bar{\ell})^1$ terms vanish
by symmetry, we have:
\bea
m^2\omega & = & +\frac{m^2}{2}\Tr[\gamma^5\slash{\epsilon}^L_a
\slash{\epsilon}^R_c 
\slash{\epsilon}^L_b ((1-z)\slash{k}+(1-y)\slash{q})
+
\gamma^5\slash{\epsilon}^L_a(-z\slash{k}-y\slash{q})
\slash{\epsilon}^L_c 
\slash{\epsilon}^R_b 
+\nonumber \\
& & \qquad \qquad 
\gamma^5\slash{\epsilon}^L_a
\slash{\epsilon}^R_c 
(-z\slash{k}+(1-y)\slash{q})
\slash{\epsilon}^R_b ]
\nonumber \\
& &
-\frac{m^2}{2}\Tr[\gamma^5\slash{\epsilon}^L_a 
\slash{\epsilon}^R_b
\slash{\epsilon}^L_c (-z\slash{k}-y\slash{q})
+
\gamma^5\slash{\epsilon}^L_a((1-z)\slash{k}+(1-y)\slash{q})
\slash{\epsilon}^L_b\slash{\epsilon}^R_c 
+\nonumber \\
& & \qquad \qquad \gamma^5\slash{\epsilon}^L_a
\slash{\epsilon}^R_b
(-z\slash{k}+(1-y)\slash{q})
\slash{\epsilon}^R_c 
]
\nonumber \\
 & = & -{m^2}\Tr[\gamma^5\slash{\epsilon}^L_a 
\slash{\epsilon}^R_b
\slash{\epsilon}^L_c (-z\slash{k}-y\slash{q})   ]
- {m^2}\Tr[\gamma^5\slash{\epsilon}^L_a 
\slash{\epsilon}^L_b
\slash{\epsilon}^R_c ((1-z)\slash{k}+(1-y)\slash{q})   ]
\nonumber \\
&  & -\;{m^2}\Tr[\gamma^5\slash{\epsilon}^L_a
\slash{\epsilon}^R_b 
\slash{\epsilon}^R_c (z\slash{k}-(1-y)\slash{q})
 ]
\eea
hence:
\bea
m^2\omega & = & -4im^2\epsilon_{\mu\nu\rho\sigma}{\epsilon}^{\mu L}_a
{\epsilon}^{\nu R}_b {\epsilon}^{\rho L}_c
(-z{k}^\sigma-y{q}^\sigma) 
-4im^2\epsilon_{\mu\nu\rho\sigma}{\epsilon}^{\mu L}_a
{\epsilon}^{\nu L}_b {\epsilon}^{\rho R}_c
((1-z){k}^\sigma+(1-y){q}^\sigma) 
\nonumber \\
& & \qquad \qquad
-4im^2\epsilon_{\mu\nu\rho\sigma}{\epsilon}^{\mu L}_a
{\epsilon}^{\nu R}_b {\epsilon}^{\rho R}_c
(z{k}^\sigma-(1-y){q}^\sigma) 
\nonumber \\
& \equiv & -4im^2\big(-z[A]^{LRL} - y[B]^{LRL}\big)
-4m^2i\big((1-z)[A]^{LLR} +(1-y)[B]^{LLR}\big)
\nonumber \\  & & \qquad 
-4m^2i\big(z[A]^{LRR} - (1-y)[B]^{LRR}\big)
\eea
We thus
have the amplitude for the pure 
$\slash{\epsilon}^{a}_L L$ vertex:
\bea
 T_L & = & (-4im^2)\times 2\int_0^1 dy \int_0^y dz\; \int \frac{d^{4}\bar\ell}{(2\pi)^{4}}
\nonumber \\ & & \left( \frac{-z[A]^{LRL} - y[B]^{LRL}
+(1-z)[A]^{LLR} +(1-y)[B]^{LLR}
+z[A]^{LRR} - (1-y)[B]^{LRR}}{
(\bar\ell^2 +z(1-z)k^2 +y(1-y)q^2 
+2k\cdot qz(1-y) - m^2  )^3} \right)
\nonumber \\
& = & -\frac{8m^2}{16\pi^2}\int_0^1 dy \int_0^y dz\;
\nonumber \\
& &  
\left(\frac{-z[A]^{LRL} - y[B]^{LRL}
+(1-z)[A]^{LLR} +(1-y)[B]^{LLR}
+z[A]^{LRR} - (1-y)[B]^{LRR}}{
2\times(m^2 +z(1-z)k^2 +y(1-y)q^2 
+2k\cdot qz(1-y)  )} \right)
\nonumber \\
& = & -\frac{1}{4\pi^2}\int_0^1 dy \int_0^y dz\;
\nonumber \\
& &  
\left(-z[A]^{LRL} - y[B]^{LRL}
+(1-z)[A]^{LLR} +(1-y)[B]^{LLR}
+z[A]^{LRR} - (1-y)[B]^{LRR}\right)
\nonumber \\
& &  
\eea
This result is negligible in the limit
$k^2, \; 2k\cdot q,\; q^2 >> m^2$. 
However, in the limit of large $m^2$
we thus have:
\bea
\label{TL}
 T_L & = & \frac{1}{24\pi^2}(
 [A]^{LRL} + 2[B]^{LRL}-2[A]^{LLR} -[B]^{LLR} -[A]^{LRR} + [B]^{LRR})
\eea
From this we can easily infer the result for a computation
of the triangle loops with a single pure  $\epsilon_\mu^{aR}\gamma^\mu R$
(right-handed) vertex. If we interchange labels
$L\leftrightarrow R$, we then  flip the overall sign, to obtain:
\bea
 T_R & = & -\frac{1}{24\pi^2}(
 [A]^{RLR} + 2[B]^{RLR}-2[A]^{RRL} -[B]^{RRL} -[A]^{RLL} + [B]^{RLL})
\eea
Combining these we have:
\bea
 T_{L} +T_{R} & = & +\frac{1}{24\pi^2}([A]^{LRL} + 2[B]^{LRL}
 -2[A]^{LLR} -[B]^{LLR} -[A]^{LRR} + [B]^{LRR}\nonumber \\
& & -[A]^{RLR} - 2[B]^{RLR}
+ 2[A]^{RRL} +[B]^{RRL} +[A]^{RLL} - [B]^{RLL})
\eea

[Now consider the application to
KK-modes following \cite{hill}. 
For  KK-mode $B_\mu^n$ we have an $x^5$ wave-function
parity of $(-1)^{n}$, and $B_{\mu L}^n = (-1)^{n}B_{\mu R}^n 
= B_{\mu}^n$. 
The KK-modes are normalized so that an axial vector
(odd $n$) couples to $\bar{\psi}\gamma_\mu \gamma^5\psi$ with positive sign.
Thus, we can write:
\bea
 T_{L} +T_{R} & = & \frac{1}{24\pi^2}((-1)^{a+c}([A] + 2[B])
 -(-1)^{a+b}(2[A] +[B]) -(-1)^{a}([A] - [B])\nonumber \\
& & -(-1)^{b}([A] + 2[B])
+ (-1)^{c}(2[A] +[B]) +(-1)^{b+c}([A] - [B])
\eea
This can be put into a compact final expression:
\bea
\label{final}
 T_{L} +T_{R} & = & \frac{1}{12\pi^2}(f_{abc}[A]^{} + g_{abc}[B])
\eea
where:
\bea
f_{abc} & = & \half((-1)^{a+c} -2(-1)^{a+b} -(-1)^{a} 
-(-1)^{b} +2(-1)^{c} +(-1)^{b+c}) \nonumber \\
g_{abc} & = & \half(2(-1)^{a+c}-(-1)^{a+b} +(-1)^{a} 
-2(-1)^{b} +(-1)^{c} -(-1)^{b+c}).
\eea
Note that  if $a + b + c $ is even, then $f = g =0$, which is
the condition that a transition cannot occur!
But, of course, 
the {\em condition that a transition can occur} is $a + b + c $ odd.
When $a + b + c $ is odd, we can therefore write:
\bea
f_{abc} & = & -(-1)^{a}-(-1)^{b} +2(-1)^{c} \nonumber \\
g_{abc} & = & (-1)^{a}-2(-1)^{b} +(-1)^{c} 
\eea
Under $b\leftrightarrow c$ we have $A\leftrightarrow -B$ and
thus $g_{abc}\leftrightarrow -f_{acb}$, which checks.
Under the Bose exchange
$a\leftrightarrow b$ we have $B\rightarrow -B$ and
$A\rightarrow A+B$ (since the $k$ in the $A$ vertex now becomes $-k-q$
with the sign flip since $a$ is incoming momentum $k+q$ and $b$ is outgoing
momentum $k$). Thus the vertex becomes:
\beq
T_{L} +T_{R}\rightarrow  \frac{1}{12\pi^2}(f_{bac}[A]^{} + (f_{bac}-g_{bac})[B])
\eeq
and we immediately verify that $f_{bac} = f_{abc}$
and $f_{bac}-g_{bac} =g_{abc}$.
Thus the amplitude is seen to be fully Bose symmetric (we leave the
verification of $a\leftrightarrow c$ Bose symmetry to the reader).

The vertex calculation can be represented by an operator of
the form:
\beq
\label{triop}
{\cal{O}} =  -\frac{1}{12\pi^2}\epsilon^{\mu\nu\rho\sigma} 
\sum_{nmk}a_{nmk}B^n_\mu B^m_\nu\partial_\rho B_\sigma^k 
\eeq
where:
\beq
a_{nmk} = \half [1 - (-1)^{n+m+k}](-1)^{m+k}
\eeq
For the process $a \rightarrow b + c$ 
the matrix element of ${\cal{O}}$ takes the form (we've multiplied
by $+i$ from $e^{iS}$):
\bea
\label{an10}
i\bra{a}{\cal{O}}\ket{b,c} & = & \frac{1}{12\pi^2} \big[
(-a_{abc}+a_{bac}+a_{bca}-a_{cba})[B]
+ (a_{acb}-a_{cab}+ a_{bca}-a_{cba})[A]
\big]
\nonumber \\
\eea
and we see that (for $a + b + c$ odd):
\bea
-a_{abc}+a_{bac}+a_{bca}-a_{cba} & = & g_{abc} \nonumber \\
a_{acb}-a_{cab}+ a_{bca}-a_{cba} & = & f_{abc}
\eea

\subsection{Massive Left-Right Symmetric Anomaly}

The current divergence, 
$\partial_\mu\bar{\psi}\gamma_\mu\psi_L$, is obtained by
the replacement $\epsilon_\mu \rightarrow p_\mu$ in $T_L$.
We thus have that
$A \rightarrow -E$ and $B\rightarrow E$:
\bea
\label{div}
\bra{0}\partial_\mu\bar{\psi}\gamma_\mu\psi_L\ket{b,c} & = & \frac{1}{24\pi^2}(
 [E]^{RL} + [E]^{LR} + 2[E]^{RR})
\eea
Likewise:
\bea
\label{divR}
\bra{0}\partial_\mu\bar{\psi}\gamma_\mu\psi_R\ket{b,c} & = & -\frac{1}{24\pi^2}(
 [E]^{LR} + [E]^{RL} + 2[E]^{LL})
\eea
This result is the current divergence
including the loop numerator mass insertions:
\bea
\label{div1L}
\partial_\mu\bar{\psi}\gamma_\mu\psi_L & = & \frac{1}{48\pi^2}(
F^L_{\mu\nu}\tilde{F}^R_{\mu\nu}+F^R_{\mu\nu}\tilde{F}^R_{\mu\nu})
\eea
\bea
\label{div1R}
\partial_\mu\bar{\psi}\gamma_\mu\psi_R & = & -\frac{1}{48\pi^2}(
F^L_{\mu\nu}\tilde{F}^R_{\mu\nu}+F^L_{\mu\nu}\tilde{F}^L_{\mu\nu})
\eea

We emphasize that this result is {\em not the anomaly}.
To extract the anomaly, we note that the equations of motion
yield the divergences of the spinor currents:
\beq
\partial^\mu \bar{\psi}\gamma_\mu \psi_L  = -im(\bar{\psi}_L\psi_R
-\bar{\psi}_R\psi_L) + \makebox{anomaly}
\eeq
\beq
\partial^\mu \bar{\psi}\gamma_\mu \psi_R  = -im(\bar{\psi}_R\psi_L
-\bar{\psi}_L\psi_R) + \makebox{anomaly}
\eeq
We thus need to subtract the vacuum to 2-gauge field 
matrix element
of the mass term, which is the operator $-im\bar{\psi}\gamma^5\psi$, 
to obtain the anomaly. The mass term
has a similar structure to the triangle diagrams, and we define:
\bea
 M^5 & = & 
 (-1)(i)^2(i)^3\int' \int \frac{d^{4}\ell}{(2\pi)^{4}}
\frac{N_1+N_2}{D} \nonumber \\
N_1 & = & (-i)(-im)\Tr[\gamma^5(\slash{\ell}-\slash{q}+m)
(\slash{\epsilon}^L_c L+\slash{\epsilon}^R_c R) (\slash{\ell}+m)
(\slash{\epsilon}^L_b L+\slash{\epsilon}^R_b R) (\slash{\ell}+\slash{{k}}+m)]
\nonumber \\
N_2 & = & (+i)(-im)\Tr[\gamma^5(\slash{\ell}+\slash{k}-m)
(\slash{\epsilon}^L_b L+\slash{\epsilon}^R_b R)(\slash{\ell}-m)
(\slash{\epsilon}^L_c L+\slash{\epsilon}^R_c R)(\slash{\ell}-\slash{q}-m)]
\nonumber \\
D &  = &( \ell^2 +2\ell\cdot(zk-(1-y)q)+zk^2 +(1-y)q^2 -m^2 )^3
\eea
The numerators become:
\bea
N_1 & = & -m\Tr[\gamma^5(\slash{\bar\ell}
-z\slash{k}-y\slash{q}+m)
(\slash{\epsilon}^L_c L+\slash{\epsilon}^R_c R) 
\times \nonumber \\ & &
(\slash{\bar\ell}-z\slash{k}+(1-y)\slash{q}+m)
(\slash{\epsilon}^L_b L+\slash{\epsilon}^R_b R) 
(\slash{\bar\ell}+(1-z)\slash{k}+(1-y)\slash{q}+m)]
\nonumber \\
N_2 & = & m\Tr[\gamma^5(\slash{\overline{\ell}}+
(1-z)\slash{k}+(1-y)\slash{q}-m)
(\slash{\epsilon}^L_b L+\slash{\epsilon}^R_b R)
\times\nonumber \\ & & 
(\slash{\bar{\ell}}-z\slash{k}+(1-y)\slash{q}-m)
(\slash{\epsilon}^L_c L+\slash{\epsilon}^R_c R) 
(\slash{\bar{\ell}}-z\slash{k}-y\slash{q}-m)]
\eea
Defining the expansion in $\bar{\ell}^2$:
\beq
N_1+N_2 = m^2 N^{(2)}\bar{\ell}^2 + m^2 \omega'\bar{\ell}^0
\eeq
we see that $N^{(2)} = 0$ since it reduces to
$\Tr(\gamma^5 \slash{a}\slash{b})=0$. Thus, the
$\omega'$ terms are:
\bea
m^2\omega' & = & 
- m\Tr[\gamma^5(-z\slash{k}-y\slash{q}+m)
(\slash{\epsilon}^L_c L+\slash{\epsilon}^R_c R) 
\times \nonumber \\ & &
(-z\slash{k}+(1-y)\slash{q}+m)
(\slash{\epsilon}^L_b L+\slash{\epsilon}^R_b R) 
((1-z)\slash{k}+(1-y)\slash{q}+m)]
\nonumber \\
&  & 
+m\Tr[\gamma^5(
(1-z)\slash{k}+(1-y)\slash{q}-m)
(\slash{\epsilon}^L_b L+\slash{\epsilon}^R_b R)
\times\nonumber \\ & & 
(-z\slash{k}+(1-y)\slash{q}-m)
(\slash{\epsilon}^L_c L+\slash{\epsilon}^R_c R) 
(-z\slash{k}-y\slash{q}-m)]
\eea
We can write:
\bea
m^2\omega' & = &
- m \Tr[(\slash{B}+m)\gamma^5(\slash{A}+m)
(\slash{\epsilon}_c )
(\slash{C}+m)
(\slash{\epsilon}_b )]
\nonumber \\
&  & + m \Tr[(\slash{A}-m)\gamma^5
(\slash{B}-m)(\slash{\epsilon}_b )
(\slash{C}-m)
(\slash{\epsilon}_c )]
\eea
where:
\bea
\slash{A} & =  & -z\slash{k}-y\slash{q} \qquad \qquad
\slash{B}  =   (1-z)\slash{k}+(1-y)\slash{q}  \qquad\qquad
\slash{C}  =   -z\slash{k}+(1-y)\slash{q}   \nonumber \\
\slash{\epsilon}_b & =  & \slash{\epsilon}^L_b L+\slash{\epsilon}^R_b R  \qquad
\qquad
\slash{\epsilon}_c  =   \slash{\epsilon}^L_c L+\slash{\epsilon}^R_c R  
\eea
\bea
m^2\omega' & = & 
-m^2
\Tr[(\slash{B}\gamma^5+ \gamma^5\slash{A})
(\slash{\epsilon}_c \slash{C}\slash{\epsilon}_b ) 
+ \slash{B}\gamma^5\slash{A}\slash{\epsilon}_c\slash{\epsilon}_b  ]
\nonumber \\
&  & 
-m^2
\Tr[(\slash{A}\gamma^5 + \gamma^5\slash{B})(\slash{\epsilon}_b \slash{C}
\slash{\epsilon}_c ) 
+ \slash{A}\gamma^5\slash{B}\slash{\epsilon}_b \slash{\epsilon}_c ]
\nonumber \\
& = & -m^2
\Tr[(\gamma^5(\slash{A} -\slash{B})[\slash{\epsilon}_c \slash{C}
\slash{\epsilon}_b - \slash{\epsilon}_b \slash{C}
\slash{\epsilon}_c] 
-\gamma^5 \slash{A}\slash{B}(\slash{\epsilon}_b \slash{\epsilon}_c 
- \slash{\epsilon}_c \slash{\epsilon}_b )]
\eea
Compute, for example, the $\epsilon^L_b\epsilon^L_c$ terms:
\bea
& = & -m^2
\Tr[\gamma^5\slash{\epsilon}^L_b\slash{\epsilon}^L_c
(\slash{A} -\slash{B})\slash{C}]
\nonumber \\
& = & 
 \; -m^2
\Tr[\gamma^5\slash{\epsilon}^L_b\slash{\epsilon}^L_c
(-\slash{k}-\slash{q} )(-z\slash{k}+(1-y)\slash{q} )]
\nonumber \\
& = & \;4i m^2 [E](1+z-y )
\eea
Full result:
\beq
m^2\omega' = 4im^2[(E^{LL} + E^{RR})(1+z-y)+ (E^{LR} + E^{RL})(y-z)]
\eeq
and:
\bea
M^5 & = & \frac{-1}{16\pi^2} \left(\frac{1}{2m^2}\right)
2\times (-4m^2)[\frac{1}{3}(E^{LL} + E^{RR})+ \frac{1}{6}(E^{LR} + E^{RL})] 
\nonumber \\
& = & \frac{1}{24\pi^2} 
[2(E^{LL} + E^{RR})+ (E^{LR} + E^{RL})] = 
\bra{ 0} -im\bar{\psi}\gamma^5\psi \ket{b,c}
\eea
or:
\bea
im\bar{\psi}\gamma^5\psi  \rightarrow
-\frac{1}{48\pi^2} 
[F_L\tilde{F}_L + F_R\tilde{F}_R + F_L\tilde{F}_R]
\eea
Forming the difference of the current
divergence with $-im\bar{\psi}\gamma^5\psi$ we have:
\bea
\label{div}
 \!\!\!\!\! \!\!\!\!\! \!\!\!\!\! \!\!\!\!\! \!\!\!\!\! \!\!\!\!\! \!\!\!\!\! \!\!\!\!\!
& & \partial_\mu\bar{\psi}\gamma_\mu\psi_L  
+  im(\bar{\psi}_L\psi_R
-\bar{\psi}_R\psi_L)
 =  \nonumber \\
 & & \!\!\!\!\! \!\!\!\!\! \!\!\!\!\! \!\!\!\!\!
\qquad \qquad  \frac{1}{48\pi^2}(
F_L\tilde{F}_R+F_R\tilde{F}_R)
-\frac{1}{48\pi^2} 
[F_L\tilde{F}_L + F_R\tilde{F}_R + F_L\tilde{F}_R]
\nonumber \\
& & =   -\frac{1}{48\pi^2} 
F_L\tilde{F}_L
\eea
Likewise:
\bea
\label{divR}
 \!\!\!\!\! \!\!\!\!\! \!\!\!\!\! \!\!\!\!\! \!\!\!\!\! \!\!\!\!\! \!\!\!\!\! \!\!\!\!\!
& & \partial_\mu\bar{\psi}\gamma_\mu\psi_R  
+  im(\bar{\psi}_R\psi_L
-\bar{\psi}_L\psi_R)
 =  \nonumber \\
 & & \!\!\!\!\! \!\!\!\!\! \!\!\!\!\! \!\!\!\!\!
\qquad \qquad  -\frac{1}{48\pi^2}(
F_R\tilde{F}_L+F_L\tilde{F}_L)
+\frac{1}{48\pi^2} 
[F_L\tilde{F}_L + F_R\tilde{F}_R + F_L\tilde{F}_R]
\nonumber \\
&   & = \frac{1}{48\pi^2} 
F_R\tilde{F}_R
\eea

\section{The Covariant Anomaly }

The consistent anomalies for the vector and axial
vector currents can be written in the form:
\bea
\partial_\mu J^\mu & = &
\frac{1}{12\pi^2}F_V^{\mu\nu} \tilde{F}_{A\mu\nu}
\nonumber \\
\partial_\mu J^{5\mu} & = &
\frac{1}{24\pi^2}[F_V^{\mu\nu} \tilde{F}_{V\mu\nu}
+F_A^{\mu\nu} \tilde{F}_{A\mu\nu}]
\eea
where $V_L = V-A$ and $V_R = V+A$ and:
\beq
J= J_L + J_R \; ,\qquad \qquad J^5 = J_R - J_L\; .
\eeq
A unique term can be added to the action of the form:
\beq
\label{WZ}
S' =  \frac{1}{6\pi^2}
\int d^4x \; \epsilon_{\mu\nu\rho\sigma}A^\mu V^\nu \partial^\rho V^\sigma
\; .
\eeq
$S'$ has even parity and is nonvanishing.
Upon variation wrt $V$ or $A$, it
adds corrections to the vector and axial currents:
\bea
\label{bardeen12}
\frac{\delta S' }{\delta V_\mu} & = & \delta J^\mu  = 
-\frac{1}{3\pi^2}\epsilon_{\mu\nu\rho\sigma}A^\nu \partial^\rho V^\sigma
+\frac{1}{6\pi^2}\epsilon_{\mu\nu\rho\sigma}V^\nu \partial^\rho A^\sigma
\nonumber \\
\frac{\delta S' }{\delta A_\mu} & = & \delta J^{5\mu}  = 
\frac{1}{6\pi^2}\epsilon_{\mu\nu\rho\sigma}V^\nu \partial^\rho V^\sigma
\eea
We see that:
\bea
\partial_\mu (\delta J^\mu) & = &
-\frac{1}{12\pi^2}\epsilon_{\mu\nu\rho\sigma}F_V^{\mu\nu} F_A^{\rho\sigma}
\nonumber \\
\partial_\mu (\delta J^{5\mu}) & = &
\frac{1}{12\pi^2}\epsilon_{\mu\nu\rho\sigma}F_V^{\mu\nu} F_V^{\rho\sigma}
\eea
The full currents, $\tilde{J} = J + \delta J$,
now satisfy:
\bea
\label{bardeen13}
\partial_\mu \tilde{J}^\mu & = & 0 \;,
\qquad \qquad \partial_\mu \tilde{J}^{5\mu}  = \frac{1}{8\pi^2}\left(
F_V^{\mu\nu}\widetilde{F}_{V\mu\nu} + \frac{1}{3}
F_A^{\mu\nu}\widetilde{F}_{A\mu\nu}
\right). 
\eea
This is called the ``{\em covariant}'' form of the anomaly.
The theory is now invariant and operators transform
covariantly with respect
to the {\em vector} gauge symmetry.

\newpage
\section{Summary}

\noindent
{\bf Pseudoscalar Mass Term}:
\bea
im\bar{\psi}\gamma^5\psi  \rightarrow
-\frac{1}{48\pi^2} 
[F_{L\mu\nu} \tilde{F}_L^{\mu\nu}+F_{R\mu\nu} \tilde{F}_R^{\mu\nu}
+F_{L\mu\nu} \tilde{F}_R^{\mu\nu}]
\eea

\noindent
{\bf Consistent Anomalies}:
\vskip 0.2in

\noindent
(1) Pure Massless Weyl Spinors ($p_i\cdot p_j >> m^2$):
\bea
\partial^\mu \bar{\psi}\gamma_\mu \psi_L  & = &
-\frac{1}{48\pi^2} F_{L\mu\nu} \tilde{F}_L^{\mu\nu}
\nonumber \\
\partial^\mu \bar{\psi}\gamma_\mu \psi_R  & = &
\frac{1}{48\pi^2} F_{R\mu\nu} \tilde{F}_R^{\mu\nu}
\eea

\noindent
(2) Heavy Massive Weyl Spinors ($p_i\cdot p_j << m^2$):
\bea
\partial^\mu \bar{\psi}\gamma_\mu \psi_L  +im(\bar{\psi}_L\psi_R
-\bar{\psi}_R\psi_L) &  = & -\frac{1}{48\pi^2} F_{L\mu\nu} \tilde{F}_L^{\mu\nu}
\nonumber \\
\partial^\mu \bar{\psi}_R\gamma_\mu \psi_R  + im(\bar{\psi}_R\psi_L
-\bar{\psi}_L\psi_R) & = & \frac{1}{48\pi^2} F_{R\mu\nu} \tilde{F}_R^{\mu\nu}
\eea

\noindent
(3) Heavy Massive Weyl Spinors ($p_i\cdot p_j << m^2$):
\bea
\partial^\mu \bar{\psi}\gamma_\mu \psi_L   &  = & 
\frac{1}{48\pi^2}(
F_{L\mu\nu} \tilde{F}_R^{\mu\nu} +
 F_{R\mu\nu} \tilde{F}_R^{\mu\nu} )
\nonumber \\
\partial^\mu \bar{\psi}\gamma_\mu \psi_R   
& = & -\frac{1}{48\pi^2} (
F_{L\mu\nu} \tilde{F}_R^{\mu\nu}
+F_{L\mu\nu} \tilde{F}_L^{\mu\nu})
\eea

\vskip 0.2in

\noindent
{\bf Consistent $L=V-A$ and $R=V+A$ Forms}:
\vskip 0.2in

\noindent
(1) Pure Massless Weyl Spinors ($p_i\cdot p_j >> m^2$):
\bea
\partial^\mu \bar{\psi}\gamma_\mu \psi  & = &
\frac{1}{12\pi^2} F_{V\mu\nu} \tilde{F}_A^{\mu\nu}
\nonumber \\
\partial^\mu \bar{\psi}\gamma_\mu\gamma^5 \psi  & = &
\frac{1}{24\pi^2}( F_{V\mu\nu} \tilde{F}_V^{\mu\nu} 
+ F_{A\mu\nu} \tilde{F}_A^{\mu\nu} )
\eea

\noindent
(2) Heavy Massive Weyl Spinors ($p_i\cdot p_j << m^2$):
\bea
\partial^\mu \bar{\psi}\gamma_\mu \psi &  = & \frac{1}{12\pi^2} F_{V\mu\nu} \tilde{F}_A^{\mu\nu}
\nonumber \\
\partial^\mu \bar{\psi}\gamma_\mu \gamma^5\psi  -2im\bar{\psi}\gamma^5\psi
& = & \frac{1}{24\pi^2}( F_{V\mu\nu} \tilde{F}_V^{\mu\nu} 
+ F_{A\mu\nu} \tilde{F}_A^{\mu\nu} )
\eea

\noindent
(3) Heavy Massive Weyl Spinors ($p_i\cdot p_j << m^2$):
\bea
\partial^\mu \bar{\psi}\gamma_\mu \psi   &  = & 
\frac{1}{12\pi^2} F_{V\mu\nu} \tilde{F}_A^{\mu\nu}
\nonumber \\
\partial^\mu \bar{\psi}\gamma_\mu \gamma^5 \psi   
& = & -\frac{1}{12\pi^2}( F_{V\mu\nu} \tilde{F}_V^{\mu\nu} )
\eea

\noindent
{\bf Covariant Forms}:
\vskip 0.2in

\noindent
Add a term to the lagrangian of
the form $(1/6\pi^2)\epsilon_{\mu\nu\rho\sigma}A^\mu V^\nu
\partial^\rho V^\sigma $. The currents are now modified to
$\tilde{J} = J +\delta J$ and $\tilde{J}^5 = J^5 + \delta J^5$
as described in the text.

\noindent
(1) Pure Massless Weyl Spinors ($p_i\cdot p_j >> m^2$):
\bea
\partial^\mu \tilde{J}_\mu   & = & 0
\nonumber \\
\partial^\mu \tilde{J}^5_\mu  & = &
\frac{1}{8\pi^2}( F_{V\mu\nu} \tilde{F}_V^{\mu\nu} 
+ \frac{1}{3}F_{A\mu\nu} \tilde{F}_A^{\mu\nu} )
\eea

\noindent
(2) Heavy Massive Weyl Spinors ($p_i\cdot p_j << m^2$):
\bea
\partial^\mu \tilde{J}_\mu &  = & 0
\nonumber \\
\partial^\mu \tilde{J}^5_\mu   - 2im\bar{\psi}\gamma^5\psi
& = & \frac{1}{8\pi^2}( F_{V\mu\nu} \tilde{F}_V^{\mu\nu} 
+\frac{1}{3} F_{A\mu\nu} \tilde{F}_A^{\mu\nu} )
\eea

\noindent
(3) Heavy Massive Weyl Spinors ($p_i\cdot p_j << m^2$):
\bea
\partial^\mu \tilde{J}_\mu   &  = & 0
\nonumber \\
\partial^\mu \tilde{J}^5_\mu  
& = & 0
\eea
The latter case is completely summarized by the fact
that, for KK-modes, 
the three-gauge boson amplitude is described by the operator:
\beq
\label{triop2}
{\cal{O}} =  -\frac{1}{12\pi^2}\epsilon^{\mu\nu\rho\sigma} 
\sum_{nmk}a_{nmk}B^n_\mu B^m_\nu\partial_\rho B_\sigma^k 
\eeq
where:
\beq
a_{nmk} = \half [1 - (-1)^{n+m+k}](-1)^{m+k}
\eeq
This operator is equivalent to 
$(-1/6\pi^2)\epsilon_{\mu\nu\rho\sigma}A^\mu V^\nu
\partial^\rho V^\sigma $
when we truncate on the first two KK-modes,
and identify $B^0 =  V$ and $B^1=A$.
Adding the
$(1/6\pi^2)\epsilon_{\mu\nu\rho\sigma}A^\mu V^\nu
\partial^\rho V^\sigma $ term cancels this quantity,
completely cancels the 3-gauge boson triangle diagrams,
and the resulting currents then have vanishing divergences.

\end{document}